\documentstyle[12pt]{article}
\begin{document}
\makeatletter
\makeatother
\begin{titlepage}
\vspace{1mm}
\begin{center}
{\Large {\bf {{Baryogenesis and phase transition in the standard model}
}}}\\
\vspace{1mm}
{\bf D.Karczewska\footnote{internet:
dkarcz@usctoux1.cto.us.edu.pl}
 R.Ma\'{n}ka\footnote{internet:
manka@usctoux1.cto.us.edu.pl} }\\
\vspace{1mm}
{\sl Department of Astrophysics and Cosmology,}\\
{\sl University of Silesia, Uniwersytecka 4, 40-007 Katowice,
Poland}\\
\end{center}
\setcounter{equation}{0}
\vspace{1mm}
\centerline{ABSTRACT}
\vspace{1mm}
The sphaleron type solution in the electroweak theory,
generalized to include the dilaton field, is examined. The solutions describe
both the variations of Higgs and gauge fields inside the
sphaleron and the  shape of the dilaton cloud surrounding the sphaleron.
Such a cloud is large and extends far outside.
These phenomena may play an important role during the baryogenesis
which probably took place in the Early Universe.\\      

\noindent
{\bf 1. INTRODUCTION}\\
In this paper the electroweak theory will be extended by the inclusion
of dilatonic field.
The Glashow-Weinberg-Salam dilatonic model with $SU_{L}(2)\times
U_{Y}(1)$
symmetry
is described by the lagrangian
\begin{eqnarray}
{\cal L} &=&  -\frac{1}{4} e^{2  \varphi (x)/f } F^{a}_{\mu
\nu}F^{a\mu \nu} -
\frac{ 1}{4 } e^{2  \varphi (x)/f } B_{\mu \nu}B^{\mu \nu}
\nonumber\\
& + & \frac{ 1}{4 }{\partial}_{\mu} \varphi {\partial}^{\mu}
\varphi + (D_{\mu}H)^{+}D^{\mu}H -U(H, \varphi)
\end{eqnarray}
with the $SU_{L}(2)$ field strength tensor $F^{a}_{\mu \nu}
= \partial_{\mu}W^{a}_{\nu} - \partial_{\nu}W^{a}_{\mu} +
g\epsilon_{abc}W^{b}_{\mu}W^{c\nu}$ and the $U_{Y}(1)$ field tensor
$B_{\mu \nu} = \partial_{\mu}B_{\nu} - \partial_{\nu}B_{\mu}$.
The covariant derivative is given by
$D_{\mu} = \partial _{\mu} -\frac{1}{2}igW^{a}_{\mu}\sigma^{a}-
\frac{1}{2}g^{'}YB_{\mu}$, where $B_{\mu}$ and 
$W_{\mu}=\frac{ 1}{2 } W^{a}_{\mu}\sigma^{ a}$
are respectively local gauge fields associated with $U_{Y}(1)$ and $SU_L(2)$
symmetry groups. $Y$ denotes the  hypercharge. The
gauge group is a simple product of $U_{Y}(1)$ and $SU_{L}(2)$
hence we have two gauge couplings $g$ and $g'$.
The generators of gauge groups are: a unit matrix for $U_{Y}(1)$ and
Pauli matrices for $SU_{L}(2)$. In the simplest version of the
standard model a doublet of Higgs fields is introduced 
$H = \left( \begin{array}{c}
H^{+} \\ H^{0} 
\end{array}   \right)
= \left( \begin{array}{c}
0 \\ \frac{ 1}{\sqrt{ 2} } v 
\end{array}   \right)$,
with the Higgs potential
\begin{equation}
U(H^{+},H, \varphi) = \lambda \left( H^{+}H - \frac{1}{2}v^{2}_0 e^{2 
\varphi (x)/f } \right)^{2} 
\end{equation}
where $f=10^7 GeV$ \cite{ro:1994} determines the dilaton scale, $v_0 =
250 GeV $, and $v$ is the vacuum  expectation value for the Higgs field. 
The form of the potential $(2)$ leads to vacuum degeneracy 
and to nonvanishing vacuum expectation value of the
Higgs field and consequently to fermion and boson
masses. In the process of spontaneous symmetry breaking the Higgs field 
acquires nonzero mass~\cite{dr:1993,mk1995}.\\

\noindent
{\bf 2. THE DILATONIC SPHALERON}\\
Let us now consider the sphaleron type solution in the electroweak theory
with dilatons. The sphaleron may be interpreted as inhomogeneous bosons
condensate $d(x),W_\mu ^a(x)$. Let us assume for simplicity that $
g'=0$. (In \cite{kb:1991} the sphaleron theory was also considered also 
for $ g' \ne 0 $.) Now we make the following anzatz 
for the sphaleron Higgs field 
\begin{equation}
H=\frac 1{\sqrt{2}}v_0U(x)h(r)\left( 
\begin{array}{c}
0 \\ 
1
\end{array}
\right) 
\end{equation}
where $U(x)=i\sum \sigma ^an^a$ and $n^a=\frac{r^a}{r}$ describe the
{\it hedgehog} structure. This produces a nontrivial topological charge of the
sphaleron. The topological charge is equal to the Chern-Simons number. Such
a {\it hedgehog} structure determines the asymptotic shape of the
sphaleron with gauge fields different from zero 
\begin{equation}
W_i^a=\epsilon _{aij}n^j\frac{1-s(r)}{gr}.
\end{equation}
Spherical symmetry is assumed for the dilaton field $d(r)$, as well as 
for the Higgs field $ h(r)$ and the gauge field $s(r)$, leading 
to the following effective lagrangian
\begin{eqnarray}
\cal{L} &=&-\frac{1}{2}f^2\left(\frac{d'(r)}{d(r)}\right)^2-\frac
{1}{2}v_0^2h'(r)^2-\frac{1}{4} \lambda v_0^4(d(r)^2-h(r)^2)^2 \nonumber\\
&-&\frac{1}{4r^2}v_0^2(3-s(r))^2h(r)^2- \frac{1}{2g^2r^2}d(r)^2\left(\frac{1}
{r^2}(3-4s(r)+s(r)^2)^2+2s'(r)^2\right)  \nonumber \\
&-&\frac{1}{4}Cv_0^4h(r)^4\left(ln(h(r)^2)-\frac{1}{2}\right) 
\end{eqnarray}
Then we switch to dimensionless variables $x=M_Wr={r}/{r_W}$,
where $M_W^2=\frac 14g^2v_0^2\sim 80GeV$, $r_W=\frac 1{M_W}\sim 10^{-18}cm$. 
The resulting Euler-Lagrange equations are following:  $s(x)$ 
function, which  describes the gauge field in the electroweak 
theory and satisfies the equation
\begin{eqnarray}
&s^{\prime \prime }(x)&+2\frac{d(x)^{\prime }}{d(x)}s^{\prime }(x)\ +\frac{%
h(x)^2}{d(x)^2}(3-s(x)) \nonumber\\
&+&\frac 1{x^2}(1-s(x))(2-s(x))(3-s(x))=0,
\end{eqnarray}
and $h(x)$ function describing the Higgs field in our theory which satisfies
the equation: 
\begin{eqnarray}
&h^{\prime \prime }(x)&+{\frac{{2\,h^{\prime }(x)}}x}+\frac 12\frac{M_H^2}{%
M_W^2}(d(x)^2-h(x)^2)h(x)\nonumber \\
&-&\frac{8C}{g^2}h(x)^3ln(h(x)^2)-\frac 1{2x^2}(s(x)-3)^2h(x)=0,
\end{eqnarray}
where $M_H^2=2\lambda v_0^2$ determines the Higgs mass.  The $d(x)$ function
describing the dependence of a dilaton field on $x$ in extended electroweak 
theory obeys the equation:
\begin{eqnarray}
&d^{\prime \prime }(x)&+\frac 2xd^{\prime }(x)-\frac{d^{\prime }(x)^2}{d(x)}+%
\frac{M_H^2}{g^2f^2}\left(\frac 1{x^4}(s(x)-3)^2(s(x)-1)^2d(x)^3\right. 
\nonumber\\
&+&\left.2\lambda {v^4}_0(d(x)^2-h(x)^2)d(x)^3+\frac 4{g^2x^2}d(x)^3s^{\prime
}(x)^2\right)=0 
\end{eqnarray}
In the last equation we have a dimensionless constant $(\frac{M_W}{gf})^
2 \sim 10^{-9}$. This practically means that the dilaton field is a free
field. The simplest solutions $h(x)=1$ (shown in Fig.1), $s(x)=3$ (Fig.2),
$d(x)=1$ (Fig.3) 
are global ones corresponding to the vacuum with broken symmetry in
the standard model.  It is obvious that far from the center of the
sphaleron our solutions should describe the normal broken phase which is
very well known from the standard model. Knowing the asymptotic solutions 
we are able to construct a two-parameter family of solutions (for details 
see \cite{mk:1996}): 
\begin{equation}
s(x)=1+2\tanh^2(tx)
\end{equation}
\begin{equation}
h(x)=\tanh(ux)
\end{equation}
\begin{equation}
d(x)=a+(1-a)\tanh^2(kx)
\end{equation}
where $t,u,a,k$ are parameters to be determined by the variational 
procedure. The relevant values of the parameters are those which 
minimize the energy. For
example, with the standard values of $M_W=80.6GeV$, $M_Z=91.16GeV$, $%
M_H=350GeV$ we found the numeric solutions $t,u,k$, as functions depending
on the initial conditions of the dilaton field $d(0)=a$ in the center of the
sphaleron. Our solutions describe both the behavior of Higgs field and
gauge field inside the sphaleron and the shape of the dilaton cloud
surrounding the sphaleron. Such a cloud is large and extends far outside the
sphaleron. The sphalerons are created during the first order phase 
transition in the expanding universe as inhomogeneous solutions 
of the motion equations.  These phase transition bubbles, 
which probably took place in the early universe, break the 
CP and C symmetry on their walls and can cause the breaking 
of baryonic symmetry. Detailed consideration of this 
problem will be the subject of a separate paper.\\

\noindent
{\bf 3. CONCLUSIONS}\\
Numerical solutions suggest that sphaleron possess an `onionlike' structure. 
In the small inner core the scalar field is decreasing with global gauge 
symmetry restoration $SU(2)\times U(1)$. In the middle layer the gauge 
field undergoes sudden change. The sphaleron coupled to dilaton field has
also an outer shell, where dilaton field changes drastically. The
spherically symmetric dilaton solutions coupled to the gauge field or
gravity are interesting in their own rights and may further influence the
monopole catalysis of baryogenesis induced by sphaleron.

This paper is sponsored by the Grant KBN 2 P304 022 06.

{\bf FIGURE CAPTIONS}\\ 
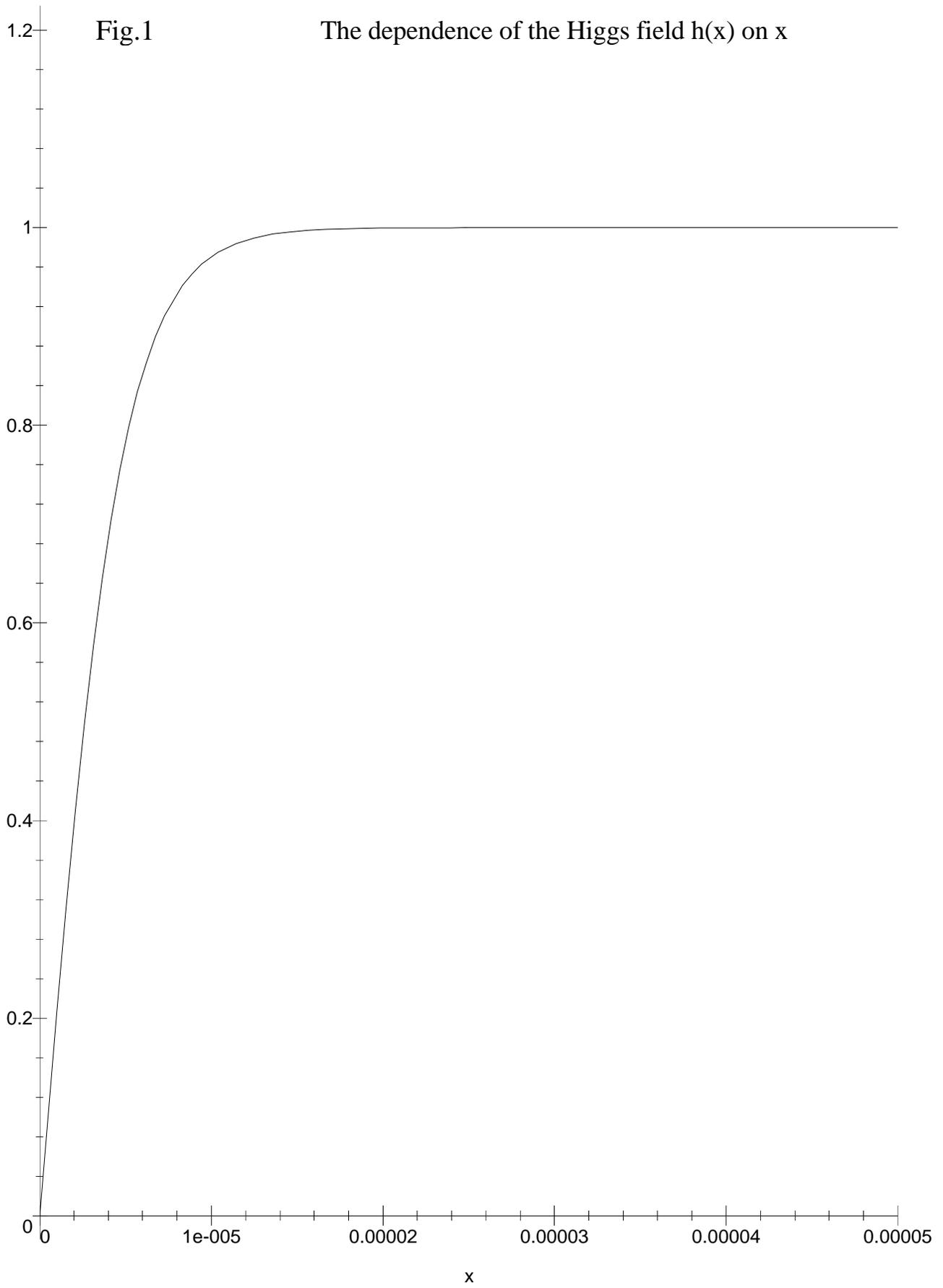
Figure 1.  The dependence of the Higgs field $h(x)$ on $x$.\\
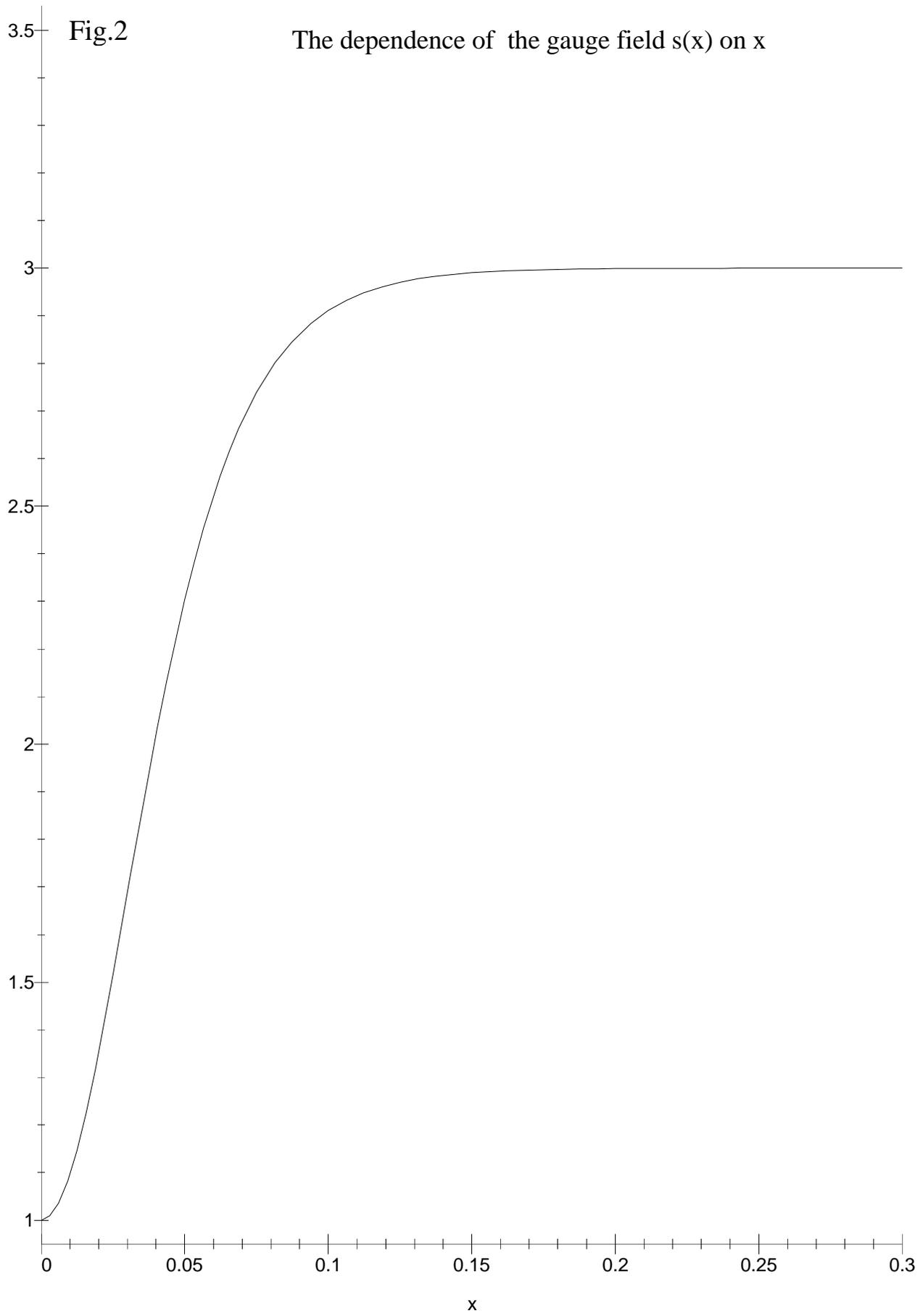
Figure 2.  The dependence of the gauge field $s(x)$ on $x$.\\
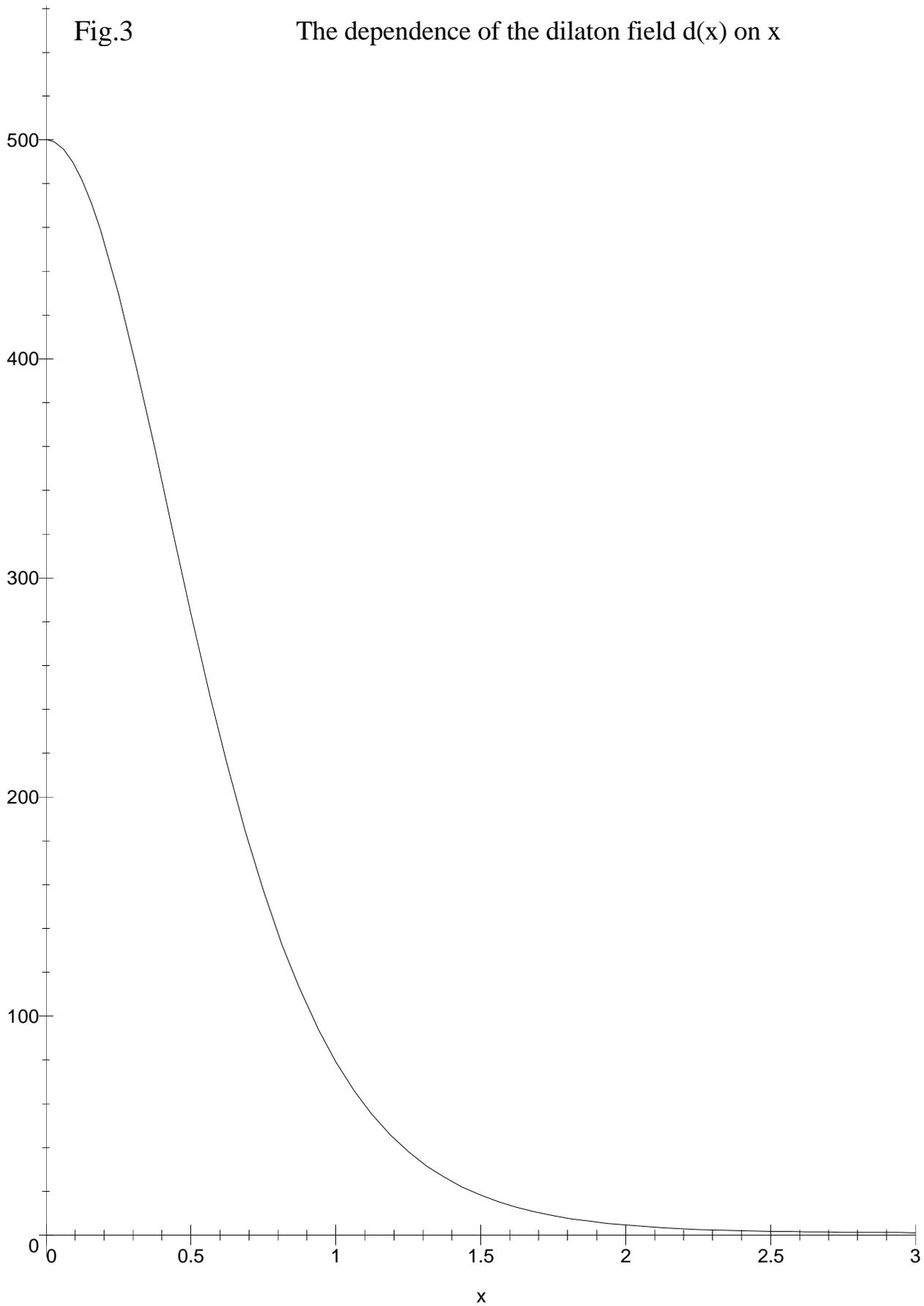
Figure 3.  The dependence of the dilaton field $d(x)$ on $x$.\\

\end{titlepage}
\end{document}